\documentclass[preprint2]{aastex}

\usepackage{graphicx}

\newcommand{\bm}[1]{\mbox{{\boldmath $#1$}}}

\newcommand{\pder}[3]{\frac{{\partial}^{#3} {#1}}{{\partial} {#2}^{#3}}}
\newcommand{\pc}{\mbox{pc} \ }

\newcommand{\Myr}{\mbox{Myr} \ }
\newcommand{\cm}{\mbox{cm} \ }

\newcommand{\km}{\mbox{km} \ }
\newcommand{\s}{\mbox{s}}

\shorttitle{Cosmic-Ray Driven Galactic Dynamo}
\shortauthors{M. Hanasz et al.}

\received{2003 December 15}
\accepted{2004 February 26}
\begin{document}

\title{Amplification of Galactic Magnetic  Fields \\
   by the Cosmic-Ray Driven Dynamo}

\author{M. Hanasz\altaffilmark{1}}
\author{G. Kowal\altaffilmark{2}, 
K. Otmianowska-Mazur\altaffilmark{2}}
\and
\author{H. Lesch\altaffilmark{3}}

\altaffiltext{1}{Centre for Astronomy, Nicholas Copernicus University,
  PL-87148 Piwnice/Toru\'n, Poland, mhanasz@astri.uni.torun.pl}
\altaffiltext{2}{Astronomical Observatory, Jagiellonian University,  
ul. Orla 171, 30-244 Krak\'ow,
kowal@oa.uj.edu.pl, otmian@oa.uj.edu.pl}
\altaffiltext{3}{Astronomical Observatory, Munich University, 
  Scheinerstr. 1, D-81679, Germany, lesch@usm.uni-muenchen.de}

\begin{abstract}
We present the first numerical model of the magnetohydrodynamical cosmic-ray
(CR) driven dynamo of the type proposed by Parker (1992). The driving force of the
amplification process comes from CRs injected into  the  galactic disk
in randomly distributed spherical regions representing  supernova  remnants.
The underlying disk is differentially rotating. An explicit resistivity is
responsible for the dissipation   of the small-scale  magnetic field component.
We obtain   amplification of  the large-scale magnetic  on a timescale 250 
Myr.
\end{abstract}


\section{Introduction}

In 1992 Parker discussed the possibility of a new kind of galactic dynamo
driven  by galactic  CRs accelerated in supernova remnants.  This 
dynamo contains a network of interacting forces: the buoyancy force of CRs,  the
Coriolis force, the differential rotation  and magnetic reconnection. Parker
estimated that such a dynamo  is able to amplify the large scale magnetic field
on timescales of the order of $10^8 {\rm yr}$.

Over the last decade we have investigated the different physical properties and
consequences  of Parker's  idea and scenario by means of analytical
calculations and numerical simulations (see eg. Hanasz \&  Lesch 1998,
Lesch \& Hanasz  2003 and references therein). Here we present the
first complete magnetohydrodynamical three-dimensional simulation  including
the full network of  relevant interacting mechanisms.

It is the aim of our contribution to show that Parker's CR driven 
dynamo indeed acts efficiently  on  timescales comparable with the disk
rotation time. In the next two Sections we  describe the physical elements of
the model and the system of equations used in numerical simulations. Section 4 
presents the numerical setup, Sections 5 and 6  inform the reader about the
results on the  structure of the  interstellar medium including CRs and
magnetic fields,  the strength of the amplified magnetic field and the  spatial
structure of the  mean magnetic field. We  summarize our results very briefly
in Section 7.

 \section{Elements of the model}
We performed computations with the aid of the Zeus-3D MHD code (Stone \& Norman
1992a,b), which we extended with  the following features: 

(1) The CR component, a relativistic gas described by the
  diffusion-advection transport equation (see Hanasz \& Lesch 2003b for the
  details of numerical algorithm). Following 
  Jokipii (1999) we presume that CRs diffuse anisotropically along
  magnetic field lines. 
(2) Localized sources of CRs:   supernova remnants, exploding randomly
   in the disk  volume (see Hanasz \& Lesch 2000). 
(3) Resistivity  of the ISM (see Hanasz et al. 2002, Hanasz \& Lesch 2003a)
   responsible for the onset of fast magnetic reconnection (in this paper we
   apply the uniform resistivity).   
(4) Shearing boundary conditions and tidal forces, following the prescription 
 by  Hawley, Gammie \&  Balbus (1995), aimed to model differentially 
 rotating  disks in the local approximation. 
(5) Realistic vertical disk gravity following the model of ISM in the Milky Way
   by Ferriere (1998).

\section{The system of equations}
We apply the following set of resistive MHD equations
\begin{equation}
\pder{\rho}{t}{}+ \bm{\nabla} \cdot (\rho \bm{V}) = 0, \label{eqofconti}
\end{equation}
\begin{equation}
\pder{e}{t}{} +\bm{\nabla}\cdot \left( e \bm{V}\right) 
    = - p \left( \bm{\nabla} \cdot \bm{V} 
\right), \label{enereq} 
\end{equation} 
\begin{eqnarray}
\pder{\bm{V}}{t}{} + (\bm{V} \cdot \bm{\nabla})\bm{V}
   = -\frac{1}{\rho} \bm{\nabla}
   \left(p + p_{\rm cr} + \frac{B^2}{8\pi}\right)   \nonumber \\    
   + \frac{\bm{B \cdot \nabla B}}{4 \pi \rho} 
      -2 \bm{\Omega} \times \bm{v}
   + 2 q {\Omega}^2 \rm x \hat{\bm{e}}_x + g_z\hat{\bm{e}}_z,
   \label{eqofmot}
\end{eqnarray}

\begin{equation}
\pder{\bm{B}}{t}{} = \bm{\nabla} \times \left( \bm{V} \times \bm{B}\right)
+ \eta \Delta \bm{B} 
\label{indeq},
\end{equation}

\begin{equation}
p=(\gamma-1) e, \quad \gamma=5/3
\end{equation}
where $\rm q=- \rm{d~ln\Omega/d~lnR}$ is the shearing parameter,  ($\rm R$ is
the distance to galactic center), $g_z$ is the vertical gravitational
acceleration, $\eta $ is the resistivity, $\gamma$ is the adiabatic index of
thermal gas,  the gradient of CR pressure  $\nabla p_{cr}$ is included
in the equation of motion 
(see Hanasz \& Lesch 2003b)
and other symbols have their usual meaning. The uniform resistivity is included
only in the induction equation (see Hanasz et al. 2002).  The adopted value
$\eta=1$ exceeds the numerical resistivity for the grid resolution defined in
the next section (see Kowal et al. 2003).   The thermal gas component is
currently treated as an adiabatic medium. 

The transport of the CR component is described by the diffusion-advection 
equation
\begin{equation}
\pder{e_{\rm cr}}{t}{} +\bm{\nabla }\left( e_{\rm cr} \bm{V}\right) 
= \bm{\nabla} \left(\hat{K} \bm{\nabla} e_{cr} \right) 
- p_{\rm cr} \left( \bm{\nabla} \cdot \bm{V} \right) 
+ Q_{\rm SN}, \label{diff-adv-eq} 
\end{equation} 
where $Q_{\rm SN}$ represents the
source term for the CR energy density: the rate of production of
CRs injected  locally in SN remnants and 
\begin{equation}
p_{\rm cr}=(\gamma_{\rm cr}-1) e_{\rm cr}, \quad \gamma_{\rm cr}=14/9.
\end{equation}
The adiabatic index of the CR gas $\gamma_{\rm cr}$ and the formula for
diffusion tensor 
\begin{equation}
K_{ij} = K_{\rm \perp} \delta_{ij} + (K_\parallel - K_{\rm \perp}) n_i n_j, 
\quad n_i = B_i/B,
\label{diftens}
\end{equation}
are adopted following the argumentation by Ryu et al. (2003).

\section{Numerical simulations}


\begin{figure*}
\includegraphics[width=0.2835\textwidth]{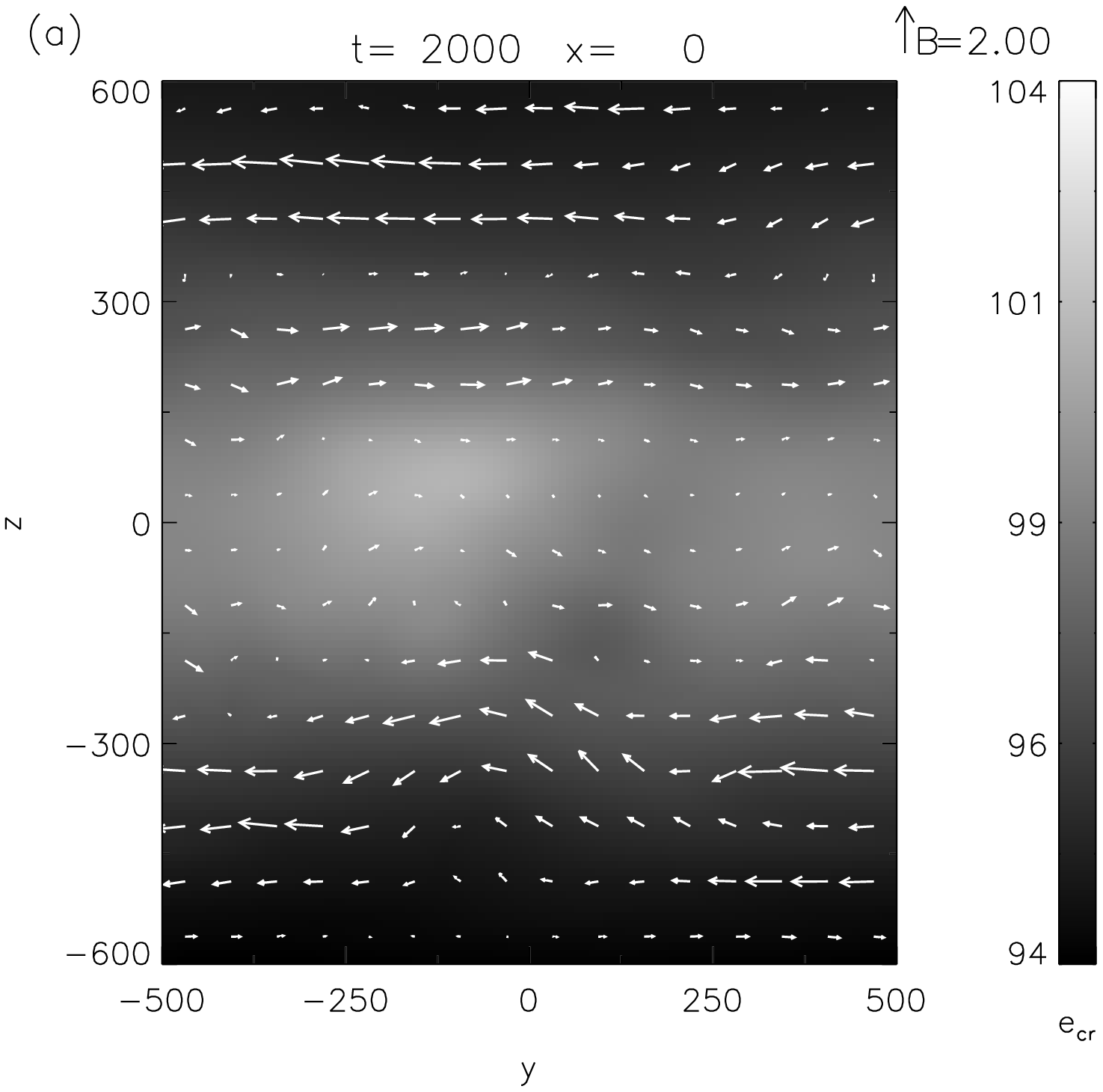}
\includegraphics[width=0.1935\textwidth]{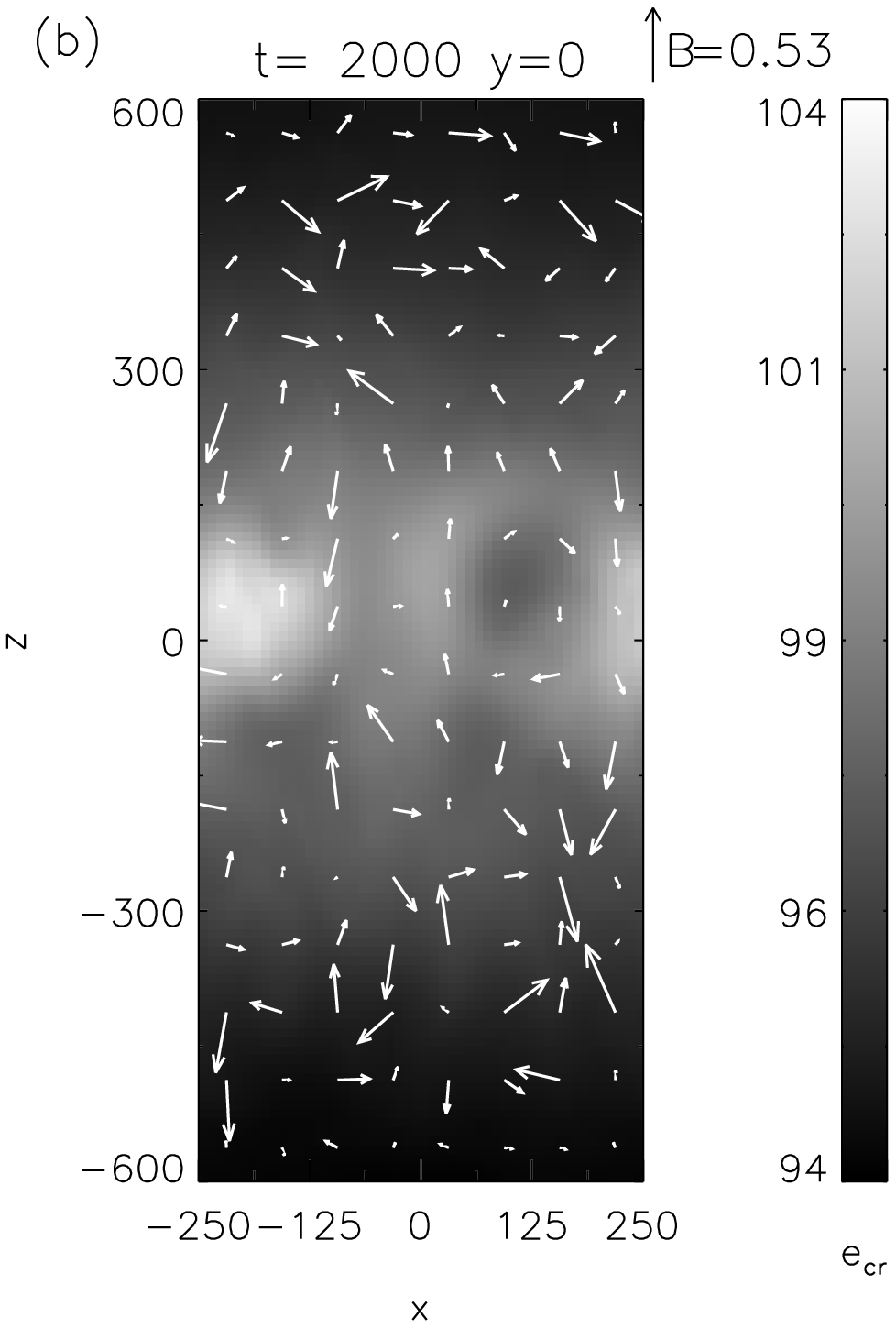}
\includegraphics[width=0.2835\textwidth]{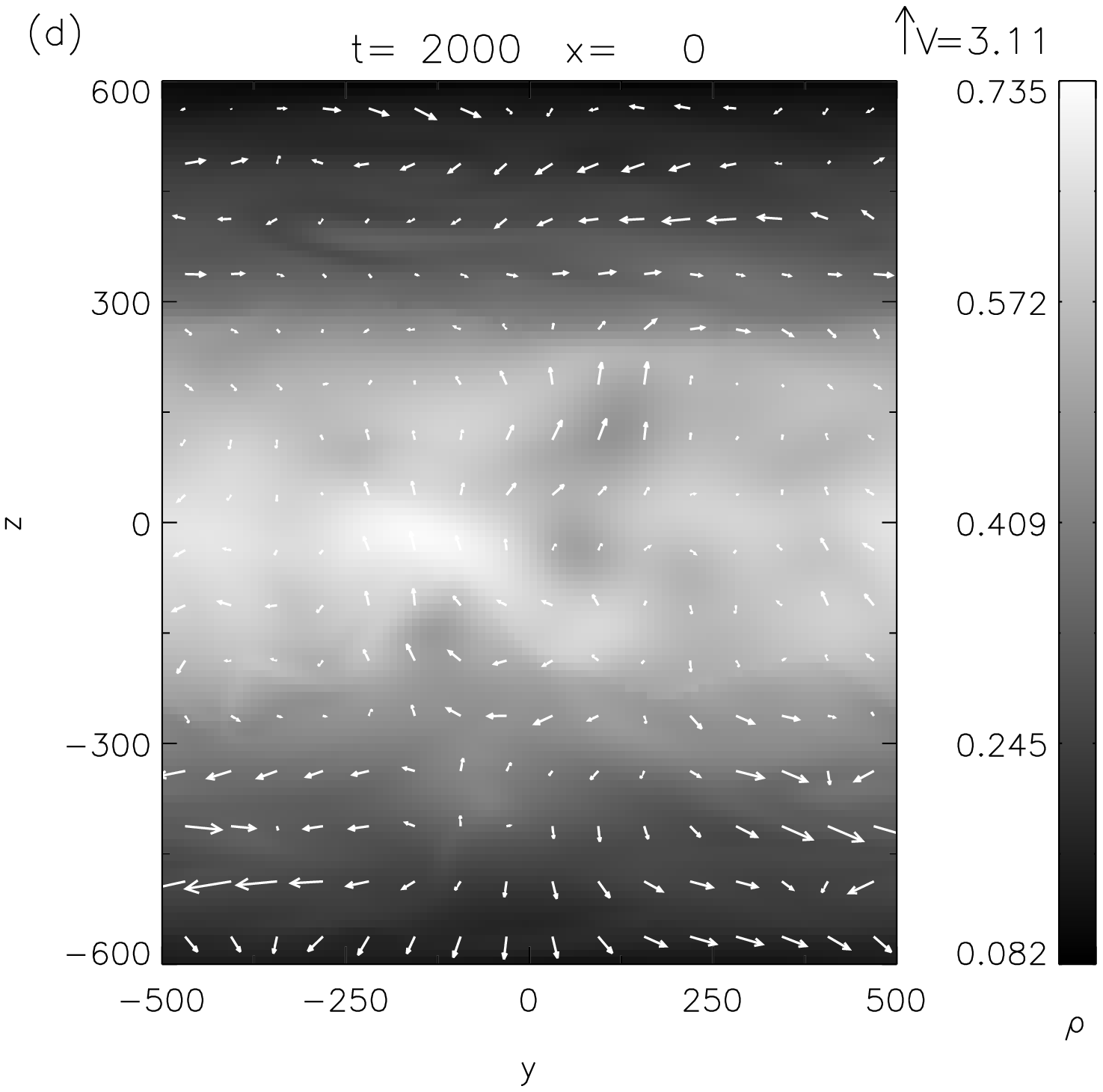}
\includegraphics[width=0.1935\textwidth]{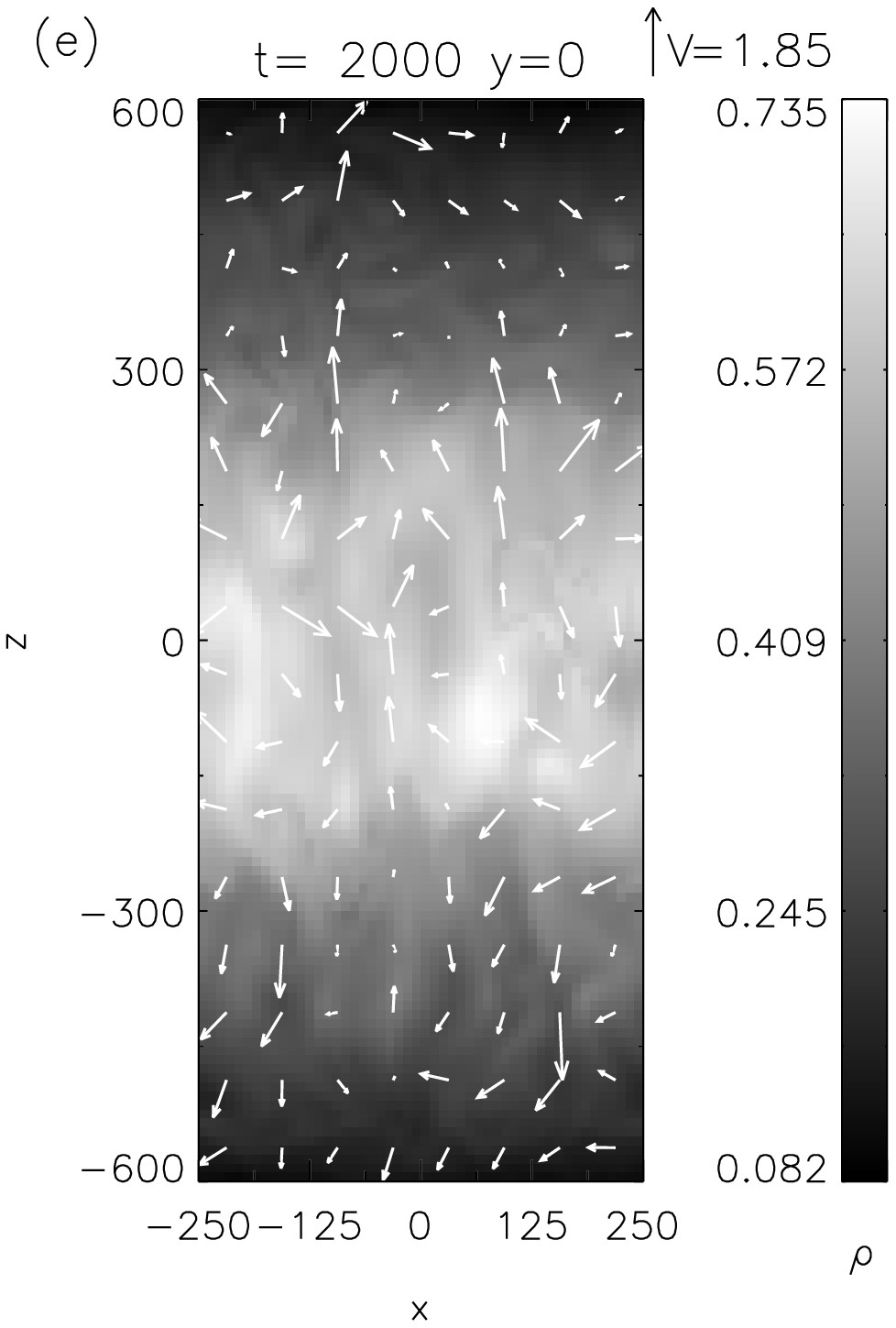}
\includegraphics[width=0.2835\textwidth]{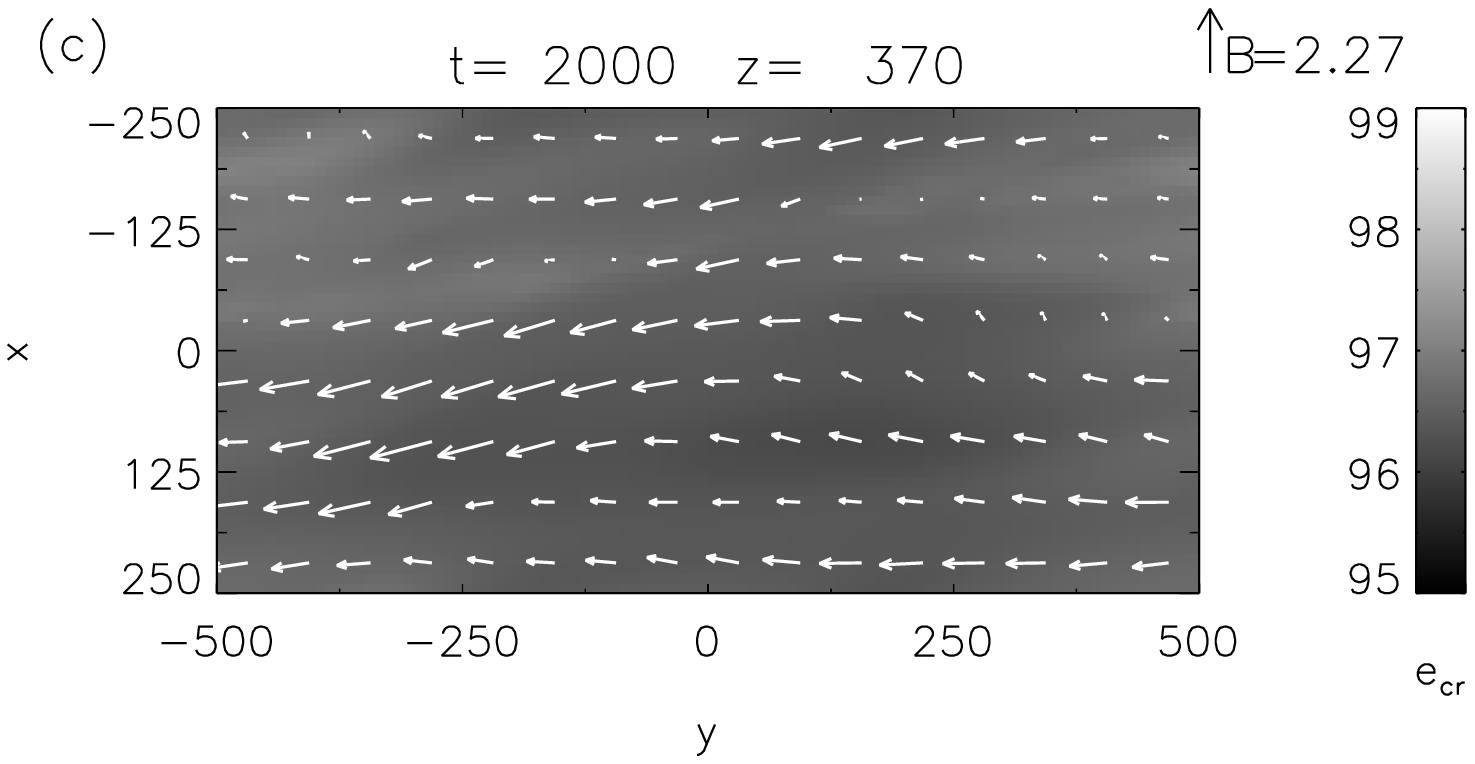}
\hspace{0.207\textwidth}
\includegraphics[width=0.2835\textwidth]{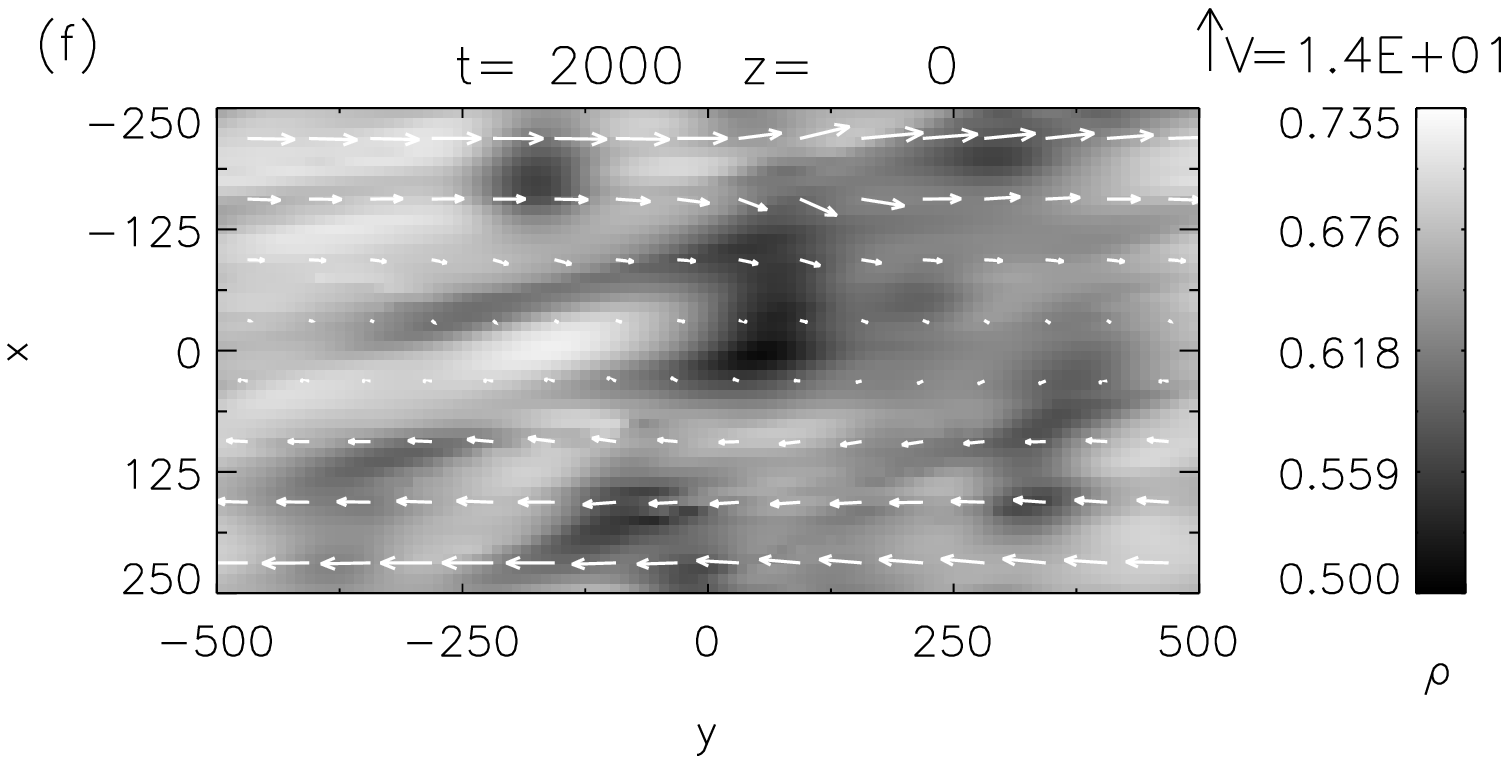}
\caption{Slices through the computational volume at $t=2000 \Myr$. Panels 
(a),(b) and (c) show CR energy density with vectors of magnetic field 
in $yz$, $xz$ and $xy$ planes respectively, panels (d), (e) and (f) show 
gas density with velocity vectors in the same planes. } 
\label{fig:slices}    
\end{figure*}

\begin{figure*}
\centerline{\includegraphics[width=0.36\textwidth]{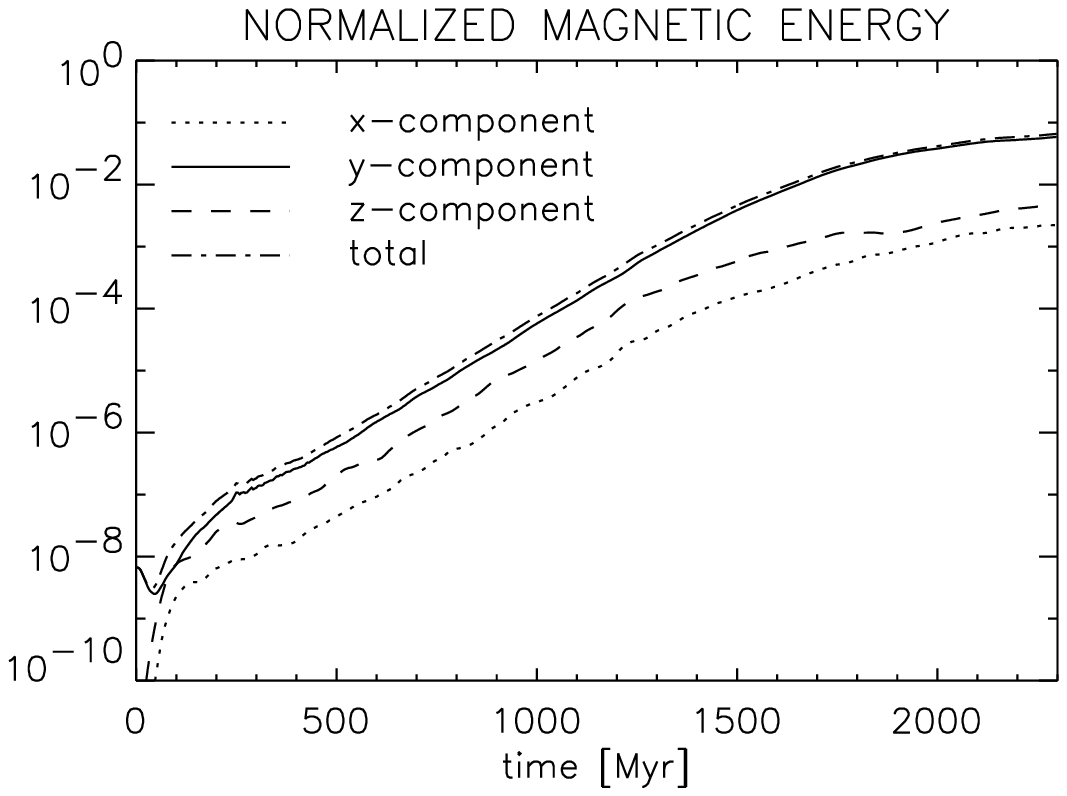}
            \includegraphics[width=0.36\textwidth]{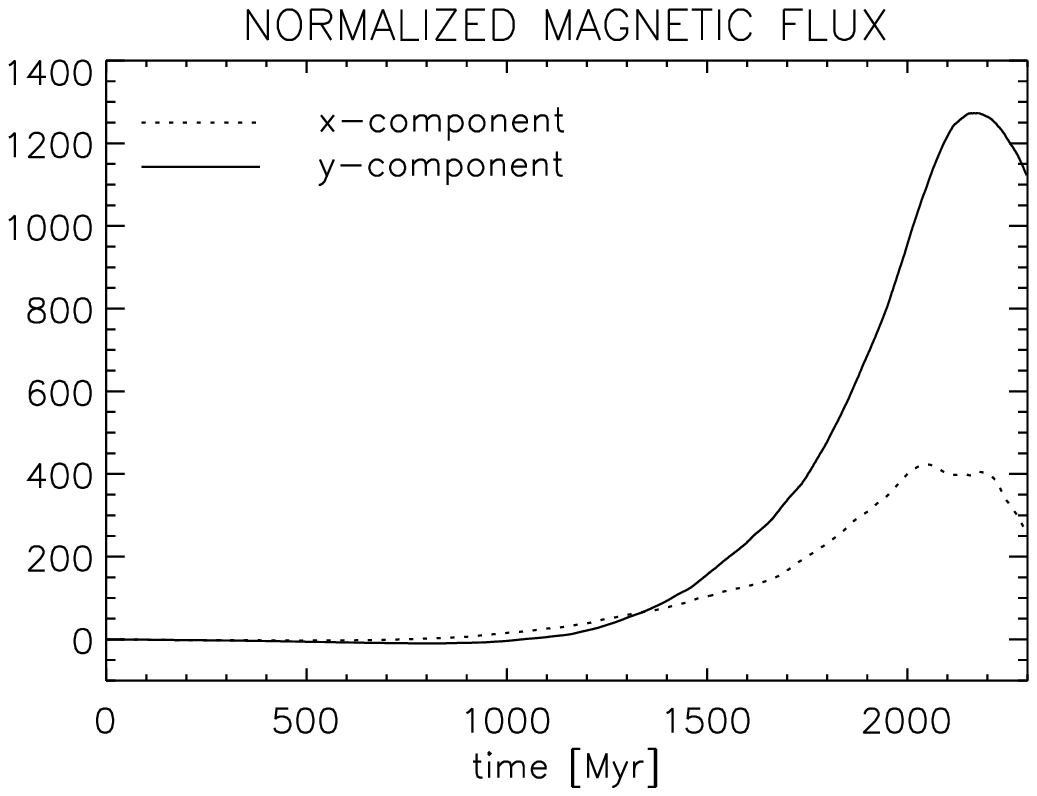}}
\caption{
Evolution of magnetic field in the computational volume. The left panel shows
the total magnetic energy normalized to the initial gas thermal energy
(dash-dot line) along with normalized energies of azimuthal magnetic field
(full line), radial magnetic field (dotted line) and vertical magnetic field  
(dashed line). The right panel shows the evolution of azimuthal magnetic flux
$\times (-1)$ 
(full line) and radial magnetic flux $\times 10$ (dotted line), both normalized
to the initial value of the azimuthal magnetic flux.}

\label{fig:amplification}
\end{figure*} 

\begin{figure*}
\centerline{
             \includegraphics[width=0.16\textwidth]{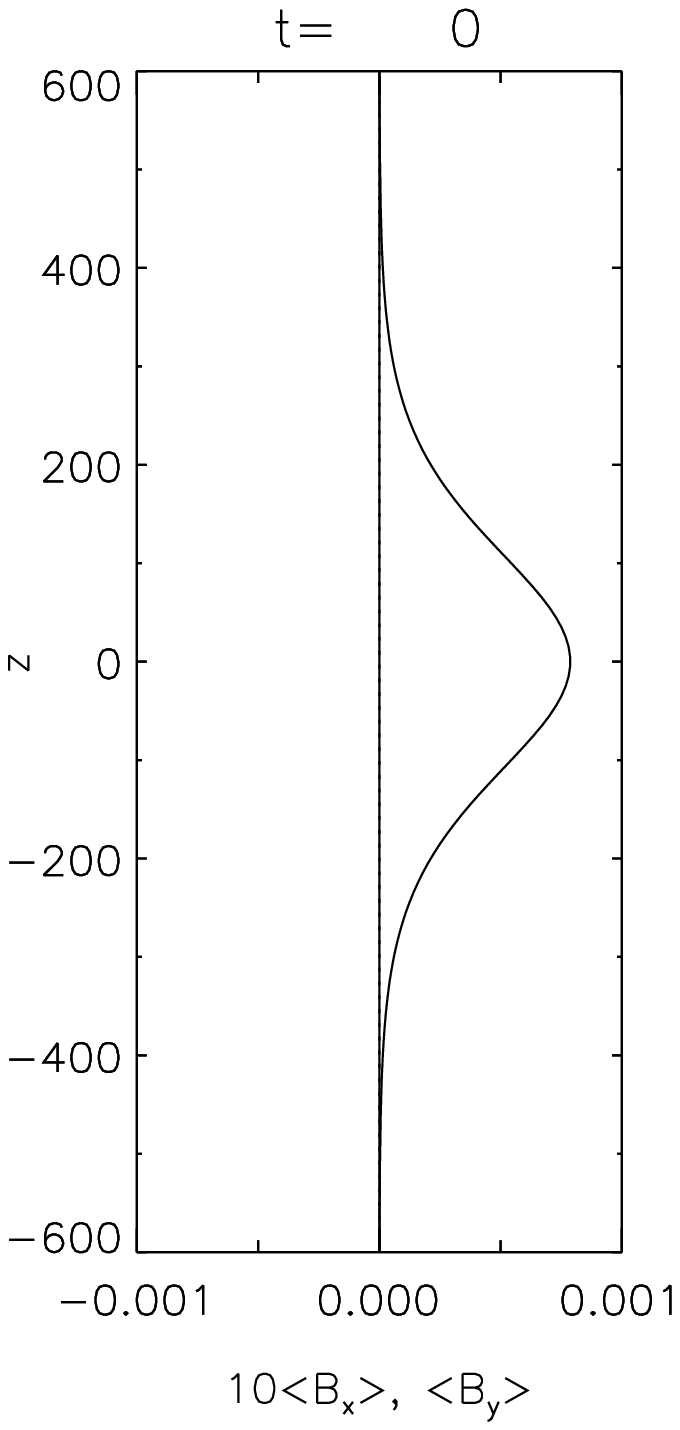}
             \includegraphics[width=0.16\textwidth]{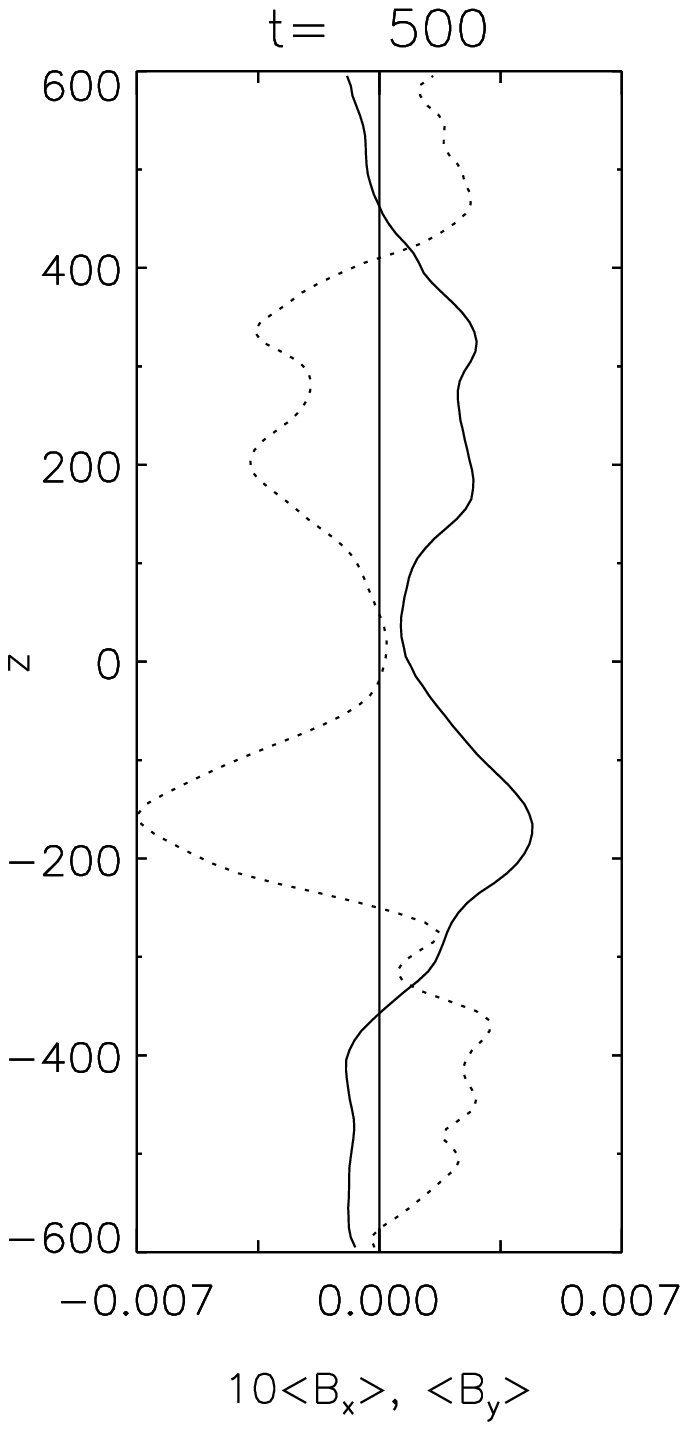}
             \includegraphics[width=0.16\textwidth]{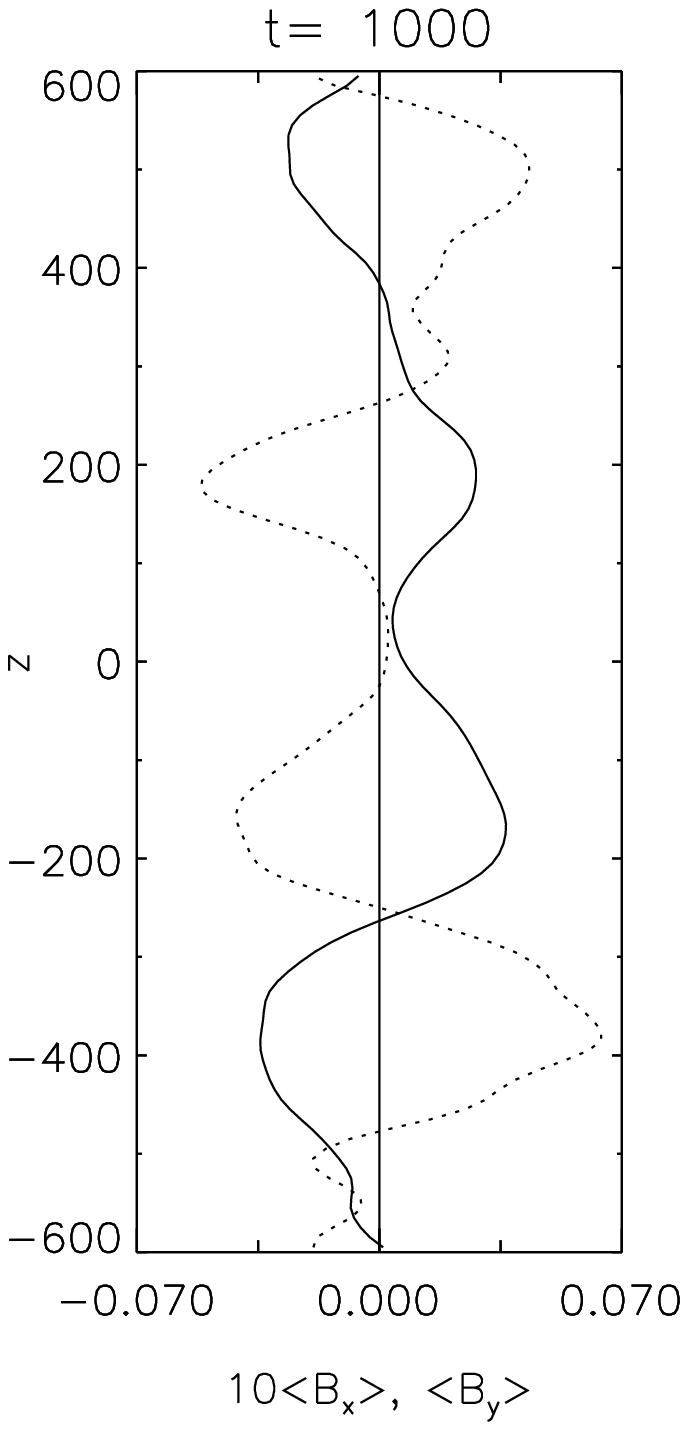}
             \includegraphics[width=0.16\textwidth]{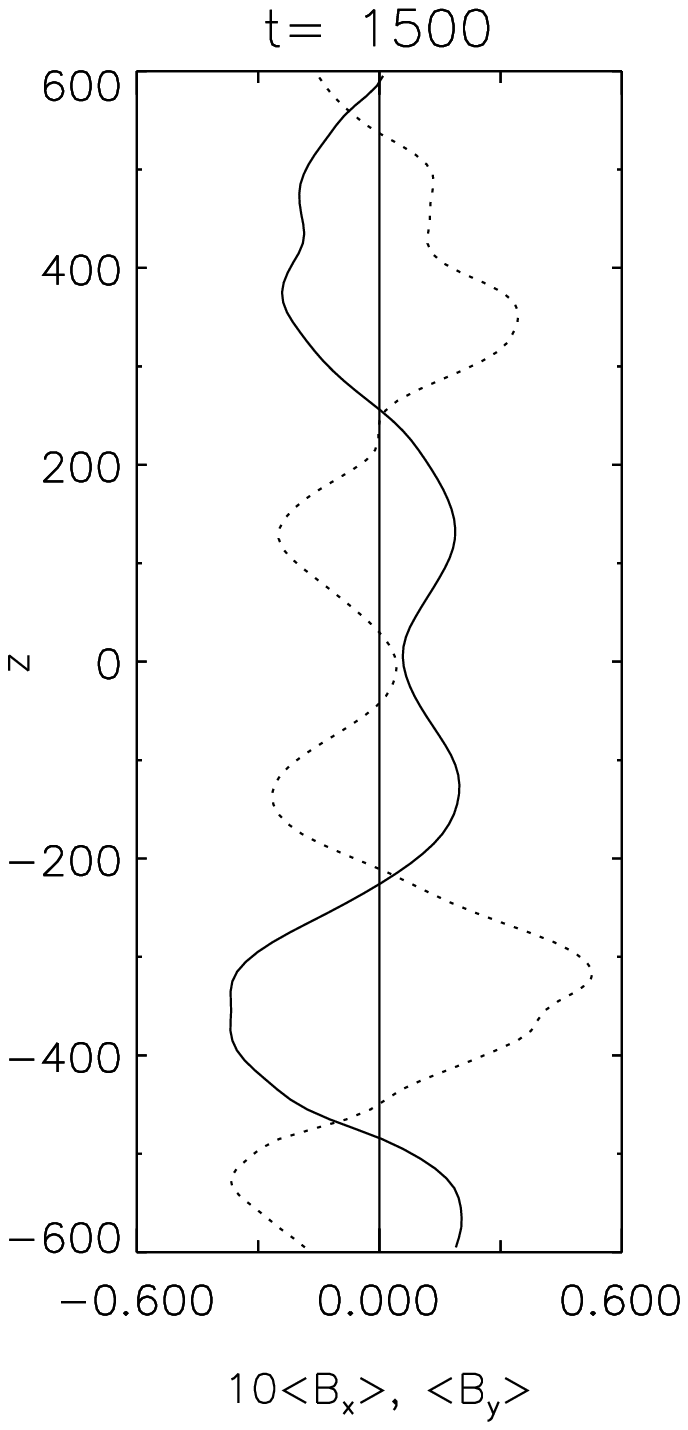}
             \includegraphics[width=0.16\textwidth]{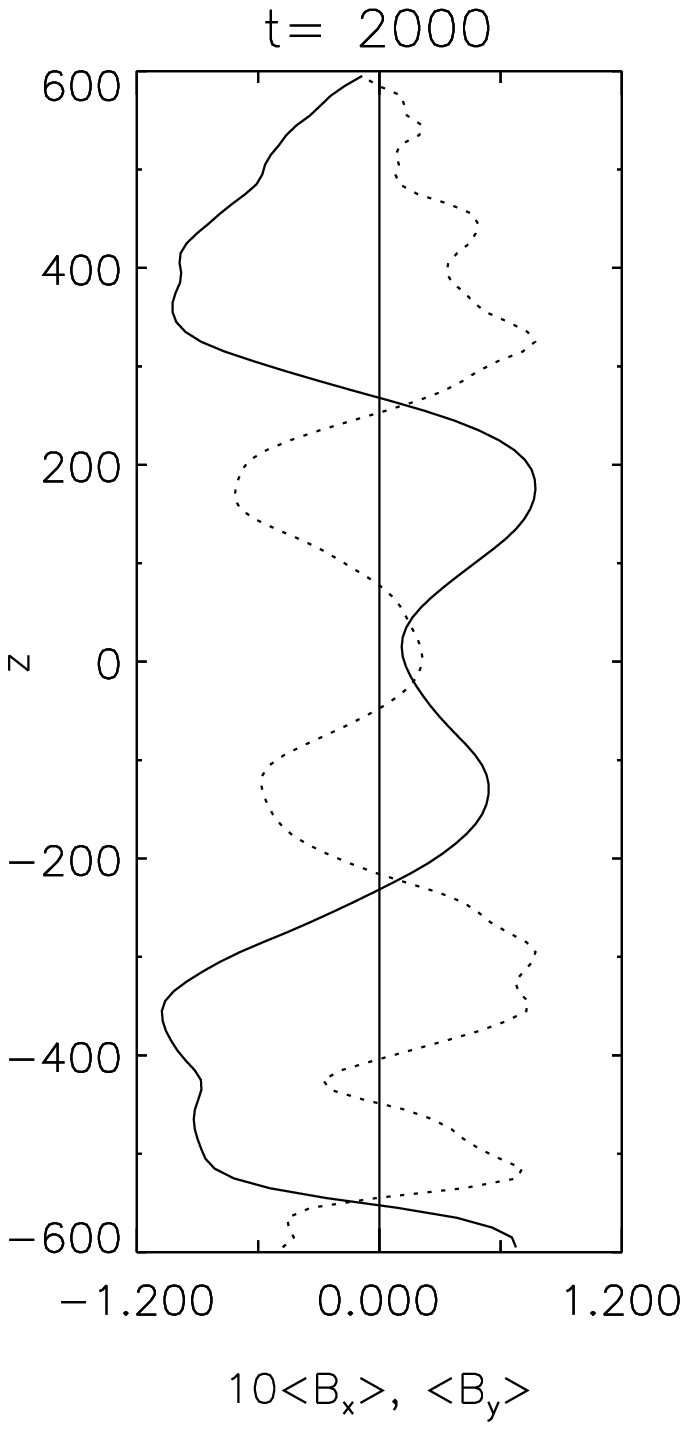} 
             \includegraphics[width=0.16\textwidth]{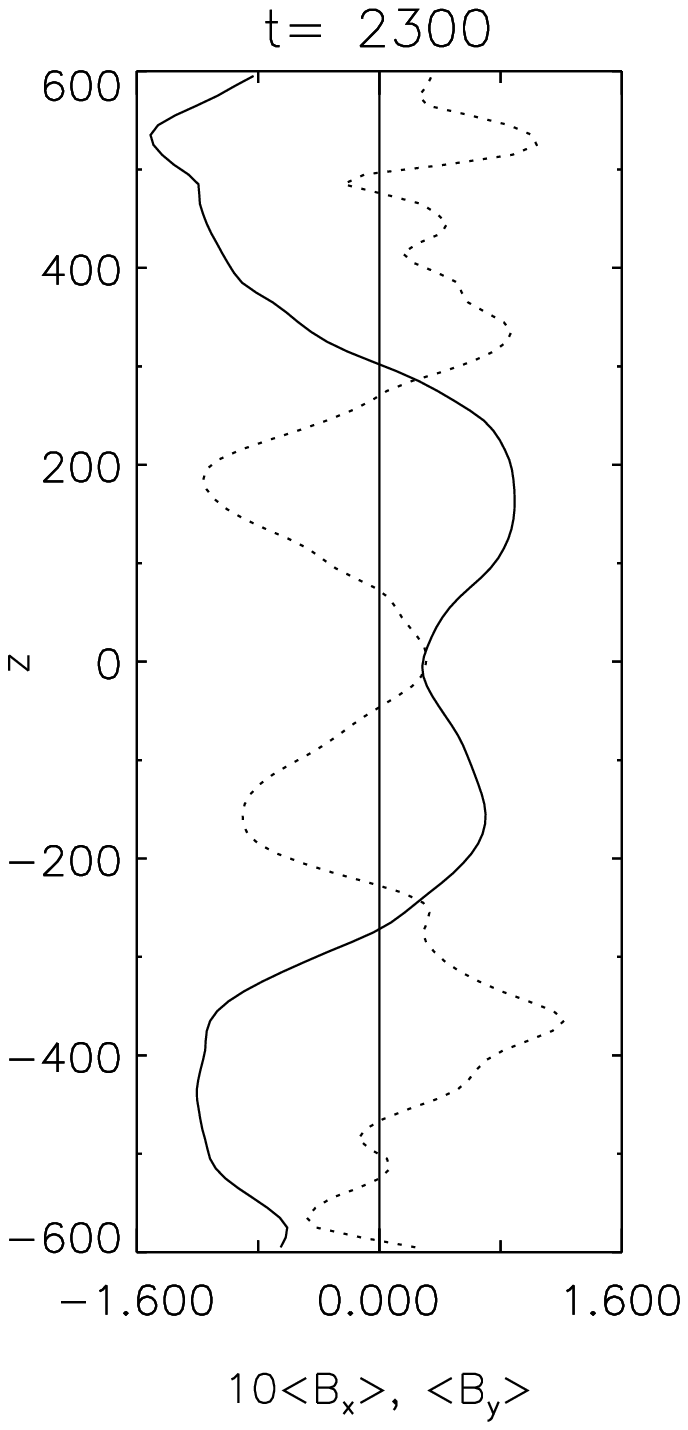}	     }
\caption{Vertical structure of the mean magnetic field for a sequence of time
instants. Spatial averaging is done over planes z=const. Full line represents
the mean azimuthal magnetic field and dotted line the radial magnetic field
multiplied by 10. The amplification effect is reflected in the varying range 
of the horizontal axis. }  
\label{fig:meanfield}   
\end{figure*} 


We performed numerical simulations in a 3D Cartesian domain  $500 \pc \times
1000 \pc \times  1200 \pc$ , extending symmetrically around the galactic 
midplane from   $z=-600 \pc$ up to $z=600 \pc$,  with the resolution of $50
\times 100 \times 120$   grid zones in directions  $x$, $y$ and $z$,
corresponding locally to cylindrical coordinates $r$, $\phi$ and $z$,
respectively.  The applied boundary conditions are periodic in the y-direction,
sheared-periodic in the $x$ direction and outflow in the $z$ direction. The
computational volume  represents a 3D region of  disk of a galaxy similar to
the Milky  Way. 

The assumed disk rotation is represented locally by the angular velocity  
$\Omega = 0.05 \Myr^{-1}$  and by a flat rotation curve corresponding to
$q=1$.    We apply the vertical gravity profile determined for the Solar
neighborhood (see  Ferriere 1998 for the formula). We assume that 
supernovae    explode with the frequency $2 {\rm kpc^{-2} \Myr^{-1}}$,   and
assume that  10 \%   of the $10^{51} {\rm erg}$ kinetic energy output from SN
is converted into the CR energy.  The CR energy is injected instantaneously
into the ISM with a Gaussian radial profile ($r_{SN}=50 {\rm pc}$) around the
explosion center.  The explosion centers are located randomly with a uniform
distribution in the $x$ and $y$ directions  and with a  Gaussian distribution
(scaleheight  $H= 100 {\rm pc}$) in the vertical direction. The applied value
of the CR parallel diffusion coefficient is  $K_\parallel = 10^4 {\rm pc^2
Myr^{-1}} = 3 \times 10^{27} {\rm \cm^2 \s^{-1}}$  (i.e. 10 \% of the realistic
value) and the perpendicular one is $K_\perp =10^3 {\rm \pc^2 \Myr^{-1}} = 3
\times 10^{26} {\rm cm^2 s^{-1}}$. 

The initial state of the system is a magnetohydrostatic  equilibrium with a
horizontal purely azimuthal magnetic field of the strength corresponding to
$p_{\rm mag}/p_{\rm gas}=10^{-8}$. The initial CR pressure in the
initial state is equal to zero.  The initial gas density at the galactic
midplane is 3 H atoms cm$^{-3}$ and the initial isothermal sound speed is
$c_{\rm si} = 7\km\s^{-1}$.  

\section{Structure of interstellar medium resulting 
from the CR-MHD simulations}

In Fig.~\ref{fig:slices} we show the distribution of CR gas together
with magnetic field, and thermal gas density together with and gas velocity in
the computational volume at $t=2000$ Myr. 

One can notice in panel (a) a dominating horizontal alignment of magnetic
vectors. The CR energy density is well smoothed by the diffusive
transport in the computational volume. The vertical gradient of the CR
energy density is maintained by the supply of CRs around the equatorial
plane in the disk in the presence of vertical gravity.  In panel (c) one can
notice that at the height $z=370\pc$  the dominating magnetic vectors are
inclined with respect to the azimuthal direction, i.e. the radial magnetic
field component is on average about 10 \% of the azimuthal one.

The CR energy density is displayed in units in which the thermal gas
energy density corresponding to $\rho=1$ and the  sound speed $c_{\rm si}= 7
\km\s^{-1}$ is equal to 1. We note that the CR energy density does not
drop to zero at the lower and upper $z$ boundaries due to our choice of outflow
boundary conditions for the CR component.
We note also that almost constant mean vertical gradient of CR energy
density is maintained during the whole simulation.

The velocity field together with the distribution of gas density,  is shown in
the next panels (d), (e) and (f). The shearing pattern of velocity can be
noticed in the horizontal slice (f). The vertical slices (d) and (e) show the
stratification of gas by the vertical gravity, acting against the vertical
gradients of thermal, CR and magnetic pressures.

\section{Amplification and structure of the mean magnetic field} 

In the following  Fig.~\ref{fig:amplification} we show how efficient is the
amplification of mean magnetic field resulting from the continuous supply of
CRs in supernova remnants.  First we note the growth of the  total
magnetic energy, by 7 orders of magnitudes during the period of 2 Gyr. Starting
from $t \sim 300$ Myr 
the growth of magnetic energy represents a straight line on a logarithmic plot,
which means that the magnetic energy grows exponentially. The e-folding time of
magnetic energy determined for the period $t=400 \div 1500$ Myr is 115 Myr.
Around  $t=1500 $ Myr, the growth starts to slow down as the magnetic energy
approaches an equipartition with the gas energy. 

The other three curves in the left panel of Fig.~\ref{fig:amplification} show
the growth  of energy of each magnetic field component. It is apparent
that the energy of radial magnetic field component is almost an order of
magnitude  smaller than the energy of vertical magnetic field component which
is almost one order of magnitude smaller than the energy of the azimuthal one.
This indicates that the dynamics of the system is dominated by the
buoyancy of CRs and that magnetic reconnection 
efficiently cancels  the excess of the random magnetic fields. 

In the right panel of Fig.~\ref{fig:amplification} we show the time evolution
of the normalized, mean magnetic fluxes $\Phi_x(t)/\Phi_y(t=0)$ and
$\Phi_y(t)/\Phi_y(t=0)$, where $\Phi_x(t)$ and $\Phi_y(t)$ are respectively
magnetic fluxes at moment $t$, threading vertical planes perpendicular to $x$
and $y$ axes respectively, and averaging is done over all possible planes of a
given type. We find that the radial magnetic flux $\Phi_x$ starts to deviate
from zero, as a result of Coriolis force and open boundary conditions.
Due to the presence of differential
rotation the azimuthal magnetic field is generated from the radial one. The
azimuthal flux grows up by a factor of 10 in the first 800 Myr of the system
evolution and then drops suddenly, reverts and continues to grow with the
opposite sign undergoing amplification by more three orders of magnitudes, with
respect to the initial value. 

In order to examine the structure of the mean magnetic field we  average of 
$B_x$ and and $B_y$  across constant z-planes. The results are presented in
Fig.~\ref{fig:meanfield} for $t=0$ (the initial magnetic field) and then for
$t=$ 500, 1000, 1500, 2000 and 2300 Myr.  We find that the mean magnetic field
grows by a factor of 10 within about 500 Myr, which gives an e-folding time
close to 250 Myr. We note that an apparent wavelike vertical structures in
$\langle B_x \rangle$ and $\langle B_y\rangle $ formed from the initial purely azimuthal,
unidirectional state  of $B_y$ and  $B_x = 0$. The evolved mean magnetic field
configuration reaches a quasisteady pattern which is  growing in magnitude with
apparent vertical reversals of  both components of the mean magnetic field. We
note also that the magnetic field at the disk midplane remains relatively
weak. 

A striking property of the mean magnetic field configuration is the almost
ideal coincidence of peaks of the oppositely directed radial and  azimuthal
field components. This feature corresponds to a picture of an
$\alpha-\Omega$-dynamo: the azimuthal mean magnetic component is generated from
the radial one and vice versa.

In order to understand better what kind of dynamo operates in our model,
we computed the $y$-component of the electromotive force $\langle{\cal E}_{{\rm
mf},y}\rangle  = \langle v_z B_x-v_x B_z\rangle $,  averaged over constant z-planes and checked
that  $\partial \langle B_x\rangle /\partial t \simeq - \partial \langle {\cal E}_{{\rm
mf},y}\rangle /\partial x$ with a reasonable accuracy.  However, we found that the
space averaged ${\cal E}_{{\rm mf},y}$ fluctuates rapidly in time, so that the
approximation of $\langle {\cal E}_{{\rm mf},y}\rangle $ by $\alpha_{yy} \langle B_y\rangle $, (where
$\alpha_{yy}$ is a component of the fluid helicity tensor), implies that
$\alpha_{yy}$ oscillates rapidly in time. This property points our model toward
the incoherent $\alpha-\Omega$ dynamo described by Vishniac \& Brandenburg
(1997). Finally, we checked that the magneto-rotational instability (Balbus and
Hawley 1991) does not seem to play a significant role in our dynamo model.  Due
to the weakness of the initial magnetic field, the wavelength of most unstable
mode of this instability remains shorter than the cell size for the first half
of the simulation time. 

\section{Conclusions}

We have described the first numerical experiment in which the effect of
amplification of the large scale galactic magnetic field was achieved by the
(1) continuous (although intermittent in space and time) supply of CRs
into the interstellar medium,  (2) shearing motions due to differential
rotation and (3) the presence of an  explicit resistivity of the medium.

We observed in our experiment the growth of magnetic energy by seven orders of
magnitude and the growth of magnetic flux by a factor of 1300 in 2150 Myr of
the system evolution. We found that the large scale magnetic field grows on a 
timescale ~ 250 Myr, which is close to the period of galactic rotation.

Therefore the galactic dynamo driven by CRs  appears to work very
efficiently, as it was suggested by Parker (1992). It is a matter of future
work to verify whether the presented model is a fast dynamo, i.e. whether it
works with a similar efficiency in the limit of vanishing 
resistivity.

\acknowledgments This work was supported by the Polish Committee for 
Scientific Research (KBN) through the grants PB  404/P03/2001/20 and PB
0249/P03/2001/21.  We thank Mordecai Marc Mac Low for the preliminary version
of shearing boundary conditions.  The presented computations have been
performed on the HYDRA computer cluster in Toru\'n Centre for Astronomy.

\end{document}